\documentclass[]{spie}  
\usepackage[]{graphicx}
\title{Science with the XEUS High Time Resolution Spectrometer} 

\author{D. Barret\supit{a}, T. Belloni\supit{b}, S. Bhattacharyya\supit{c}, E. Cackett\supit{d}, M. Gilfanov\supit{e}, E. {G\"o\u{g}\"u\c{s}}\supit{f}, J. Homan\supit{g}, M. M\'endez\supit{h}, J. M. Miller\supit{i}, M. C. Miller\supit{j}, S. Mereghetti\supit{k}, S. Paltani\supit{l}, J. Poutanen\supit{m}, J. Wilms\supit{n}, A. A. Zdziarski\supit{o}
\skiplinehalf
\supit{a}Centre d'Etude Spatiale des Rayonnements, CNRS-UPS-OMP, 9 Avenue du Colonel Roche, 31028 Toulouse Cedex 04, France; \\
\supit{b}INAF - Osservatorio Astronomico di Brera, Via E. Bianchi 46, I-23807 Merate, Italy; \\
\supit{c} Department of Astronomy and Astrophysics, Tata Institute of Fundamental Research, Mumbai, India; \\
\supit{d} University of Michigan, Department of Astronomy, 500 Church St., Ann Arbor, MI 48109-1042 ; \\
\supit{e} Max-Planck-Institute fur Astrophysik, Karl-Schwarzschild-Str. 1, D-85740 Garching bei Munchen, Germany, and Space Research Institute, Russian Academy of Sciences, Profsoyuznaya 84/32, 117997 Moscow, Russia; \\
\supit{f} Sabanc{\i} University, Faculty of Engineering \& Natural Sciences, Orhanl{\i}$-$Tuzla 34956 {\.I}stanbul, Turkey; \\
\supit{g} MIT Kavli Institute for Astrophysics and Space Research, Cambridge, MA, USA; \\
\supit{h} Kapteyn Astronomical Institute, University of Groningen P.O. Box 800, 9700 AV Groningen, The Netherlands; \\
\supit{i}Department of Astronomy, University of Michigan, 500 Church Street, Ann Arbor, MI 48109, USA; \\
\supit{j} University of Maryland, Department of Astronomy, College Park, MD 20742-2421, USA; \\
\supit{k}INAF - Istituto di Astrofisica Spaziale  e Fisica Cosmica -  via Bassini 15, I-20133 Milano, Italy; \\
\supit{l}ISDC Data Centre for Astrophysics, Geneva Observatory, Ch. d'ƒcogia 16, Ch-1290 Versoix, Switzerland; \\
\supit{m}Astronomy Division, Dept. of Physical Sciences, PO Box 3000, 90014 University of Oulu, Finland; \\
\supit{n} Dr.\ Remeis-Observatory, University of Erlangen-Nuremberg, Sternwartstr. 7, 97049 Bamberg, Germany
\& ECAP, Erwin-Rommel-Str. 1, 91058 Erlangen, Germany; \\
\supit{o}N. Copernicus Astronomical Center, Bartycka 18, 00-716 Warszawa, Poland;\\
}

\authorinfo{Further author information: (Send correspondence to Didier Barret)\\Didier Barret: E-mail: Didier.Barret@cesr.fr}

\begin{document} 
 \maketitle 

\begin{abstract}
XEUS has been recently selected by ESA for an assessment study. XEUS is a large mission candidate for the Cosmic Vision program, aiming for a launch date as early as 2018. XEUS is a follow-on to ESA's Cornerstone X-Ray Spectroscopy Mission (XMM--Newton). It will be placed in a halo orbit at L2, by a single Ariane 5 ECA, and comprises two spacecrafts. The Silicon pore optics assembly of XEUS is contained in the mirror spacecraft while the focal plane instruments are contained in the detector spacecraft, which is maintained at the focus of the mirror by formation flying. The main requirements for XEUS are to provide a focused beam of X-rays with an effective aperture of 5 m$^2$ at 1 keV, 2 m$^2$ at 7 keV, a spatial resolution better than 5 arcsec, a spectral resolution ranging from 2 to 6 eV in the 0.1--8 keV energy band, a total energy bandpass of 0.1--40 keV, ultra-fast timing, and finally polarimetric capabilities. The High Time Resolution Spectrometer (HTRS) is one of the five focal plane instruments, which comprises also a wide field imager, a hard X-ray imager, a cryogenic spectrometer, and a polarimeter. The HTRS is unique in its ability to cope with extremely high count rates (up to 2 Mcts/s), while providing sub-millisecond time resolution and good (CCD like) energy resolution. In this paper, we focus on the specific scientific objectives to be pursued with the HTRS: they are all centered around the key theme "Matter under extreme conditions" of the Cosmic Vision science program. We demonstrate the potential of the HTRS observations to probe strong gravity and matter at supra-nuclear densities. We conclude this paper by describing the current implementation of the HTRS in the XEUS focal plane.\end{abstract}



\section{INTRODUCTION}
\label{sec:intro}  

As stated in the executive summary of the Cosmic Vision proposal to ESA\footnote{http://www.xray.mpe.mpg.de/$~$xeus/}, the X-ray Evolving Universe Spectroscopy mission, XEUS, is Europe's next generation X-ray observatory\cite{hasinger}, designed to address two of the four main questions posed in Cosmic Vision, namely {\it What are the fundamental laws of the Universe?} and, {\it How did the Universe originate, and what is it made of?} XEUS will be the observatory best suited to tackle at least three of the twelve major topics laid down in Cosmic Vision: {\it The evolving violent Universe, The Universe taking shape, and Matter under extreme conditions}. With unprecedented sensitivity to the hot, million-degree Universe, XEUS will provide the long-sought answers to key questions in contemporary astrophysics:
\begin{itemize}
\item How did supermassive black holes form and grow?
\item How did feedback from these black holes influence galaxy growth?
\item How did large scale structure evolve?
\item How did the baryonic component of this structure become chemically enriched?
\item How does gravity behave in the strong field limit?
\item What makes up the core of neutron stars?
\end{itemize}
In this paper, we focus on the last two items, emphasizing on the capabilities required by XEUS to address them using bright galactic X-ray sources.
\section{MATTER UNDER EXTREME CONDITIONS}
The most extreme physical conditions in the observable Universe, regions with strongest gravity, highest densities, hottest temperatures and largest magnetic fields, occur around neutron stars (NS) and black holes (BH). They are responsible for the most dramatic events and powerful sources known, and test physics and astrophysics to the limit. Understanding how strong gravity works and testing our understanding of General Relativity (GR) requires observations of matter and radiation in regions just outside the event horizon of BH or at the surface of NS. The extreme gravity there produces large Doppler shifts, gravitational redshifts and light bending, as well as frame dragging if the central mass is rotating. Studies of the accretion flow around and upon the surfaces of NS provide constraints on the strong interaction and thus the equation of state of dense matter. All these effects can be seen or inferred from the X-ray spectra and variability of BH and NS, since X-rays are a major component of the radiation from the innermost parts of accretion flows and compact surfaces. XEUS will address major open questions like:
\begin{itemize}
\item How does gravity in the strong field limit work and what are the properties of curved spacetime?
\item What is the equation of state of dense matter?
\item What is the geometry of accretion flows around black holes and neutron stars?
\end{itemize}

\subsection{Strong gravity}
The energy spectrum of accreting compact objects displays several components, all varying in response to a changing accretion rate. In the most common spectral state, the accretion geometry consists of an optically thick accretion disc generating quasi-blackbody radiation with a coronal region above and below it\cite{done}. Comptonization in the corona of soft photons from the disc produces a power-law like X-ray continuum, irradiating the disk and giving rise to a reflection spectrum, consisting of backscattered continuum with superposed fluorescent and recombination lines. Such a reflection component has been detected in a wide range of accreting systems, from active galactic nuclei (AGN) to NS low mass X-ray binaries\cite{miller}. In those systems, the whole reflection spectrum is relativistically broadened, in particular the Iron K$\alpha$ fluorescence line. The K$\alpha$ line shape is distorted by various relativistic effects such as gravitational redshift, light-bending, frame-dragging, and Doppler shifts, with the effects becoming stronger close to the central object. Measurements of the degree of broadening then translate into values of the innermost radius of the disc. Through the effects of frame dragging on the innermost stable orbit, this leads to the determination of the spin of the BH\cite{miller,reisgx339}. Similarly for NS, any measurements of the inner disk radius limit the NS radius, hence the equation of state of dense matter\cite{cackett,bhat2007}. The properties of the line and the reflection component constrain the properties of the accretion flow, within a few times the radius of the event horizon, in a domain where GR is pushed to the strong field limit. 

Although supermassive BH in AGN produce more line photons per orbital cycle than stellar mass compact objects, the statistical properties of the line in the latter systems can still be inferred from time averaged spectra. In addition, compared to AGN, galactic systems have the advantage to vary on much shorter timescales further displaying large luminosity changes (by several orders of magnitude in the case of transient sources). This provides the opportunity to probe accretion under very different conditions. For instance, inferring a consistent BH spin through accurate modeling of energy spectra observed in different luminosity states would demonstrate the validity of the assumptions made on the underlying physical processes, including GR. 

\begin{figure}
\begin{center}
\begin{tabular}{c}
\includegraphics[height=7cm]{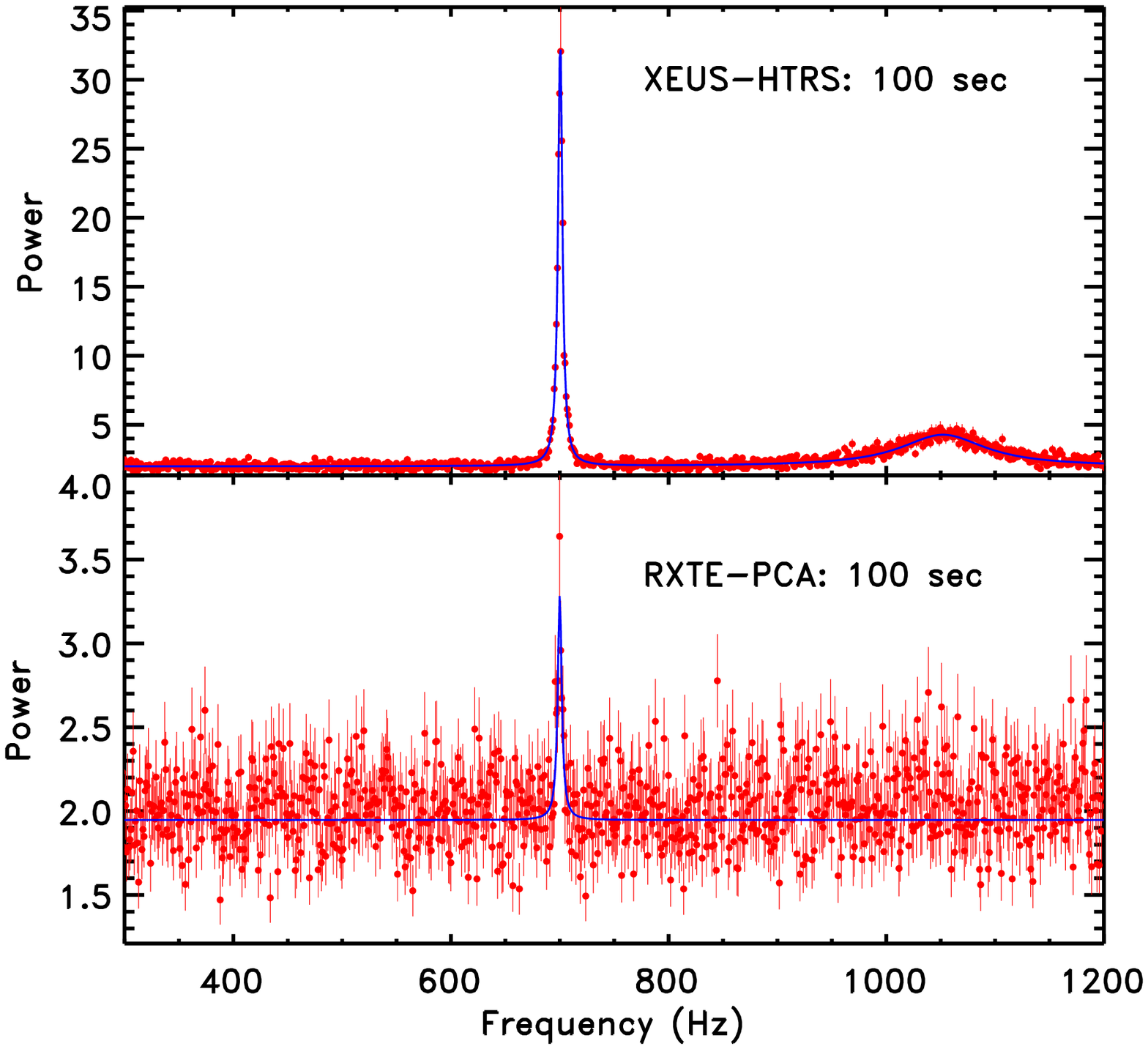}\includegraphics[height=7cm]{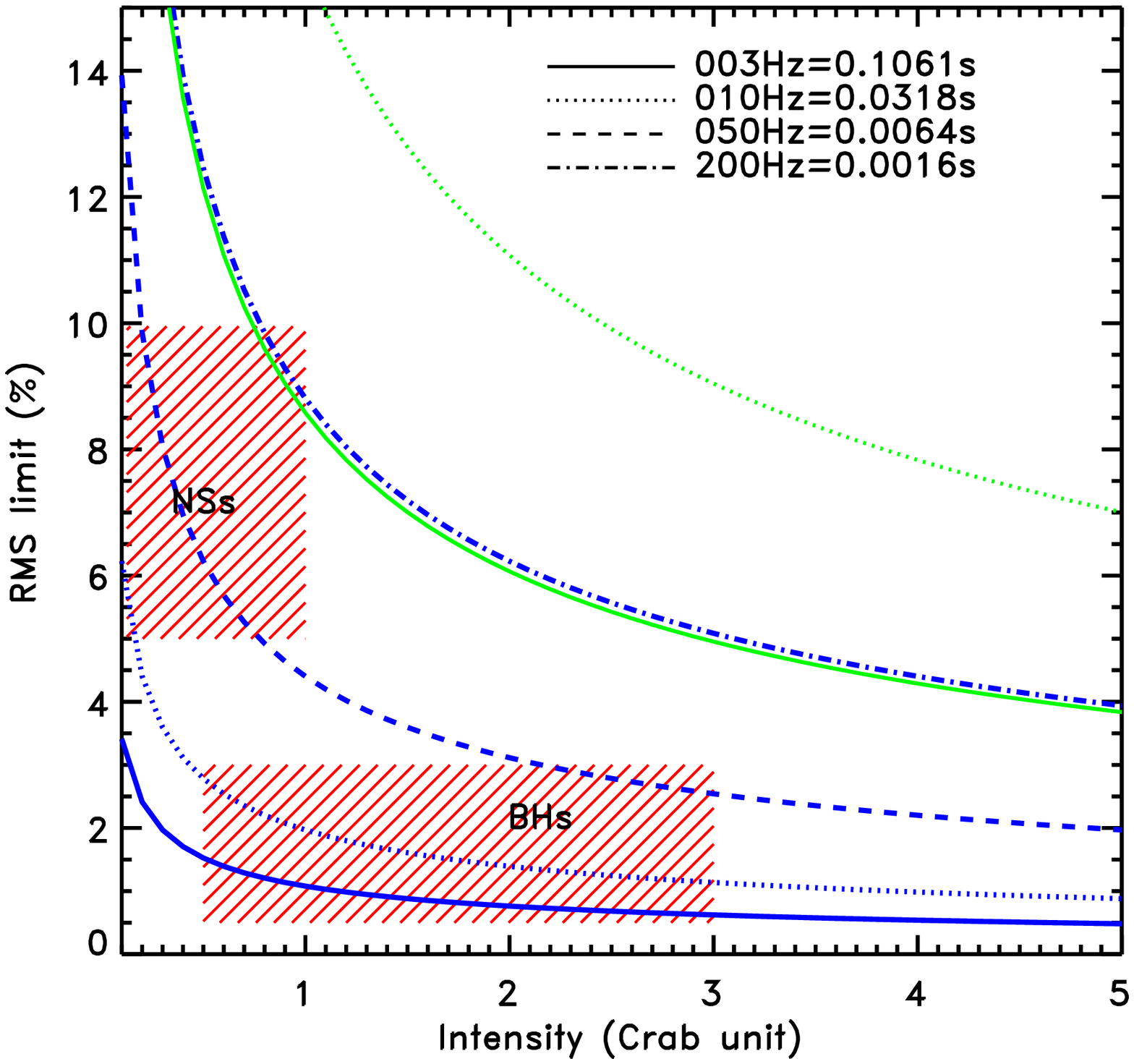}
\end{tabular}
\end{center}
\caption[example] 
{ \label{fig:cohedet} 
Left) A comparison of two Power Density Spectra, as recorded by the HTRS and by the RXTE/PCA. The gain in sensitivity is obvious, allowing to detect the lower and the upper kHz QPOs on comparable timescales. Right) In blue, the RMS amplitude limit to detect QPOs on their coherence timescales for various QPO widths (the coherence time is given as $1/\pi\Delta\nu$), as a function of source count rate normalized to the intensity of the Crab (equivalent to a count rate of 457000 counts/s above 0.2 keV). The regions where NS and BH QPOs fall are represented as red-dashed rectangles. The current limit obtained with the PCA are shown as green lines, for a width of 3 Hz (solid green line) and 10 Hz (dotted green line).  }
   \end{figure}

In a way complementary to X-ray spectroscopy, X-ray timing also probes accretion disk physics on the dynamical timescales of the inner parts of the flow, providing independent measurements of the same quantities inferred from the Iron line properties\cite{miller05}. This is particularly the case for high frequency QPOs detected by the Rossi X-ray Timing Explorer\cite{vdk}. There are strong evidences that X-ray QPOs are associated with the orbital motion of matter under very extreme conditions of gravity, temperatures (up to 10$^9$ K) and velocities (0.5c). Although differing in the details of the interpretation, most models of associate QPOs with general relativistic (GR) frequencies, which then depend on the spin and mass of the central compact object. Interestingly enough it has also been proposed that high frequency QPOs may reveal strong field GR signatures, such as the existence of an innermost stable orbit\cite{barret}. Similarly, using their mass scaling properties, QPOs have been claimed as a tool to infer the mass of the suspected intermediate mass BH in the puzzling ultra-luminous X-ray sources\cite{abra04} (QPOs may also be used to estimate the mass of the BH in some AGN). Breakthroughs in understanding QPOs will arise from their observations on timescales closer to the coherence time of the underlying oscillator (see Fig. \ref{fig:cohedet}), from the detection of the weakest features (e.g. sidebands\cite{jonker}) to observe additional modes of oscillation and to identify unambiguously QPO frequencies. At the same time, one can expect that in the XEUS timeframe theoretical understanding of accretion disk physics as well as global disk simulations will advance to provide the necessary framework for exploiting the potential of QPOs for probing GR in the strong field regime, and constraining the mass, the radius, the spin of compact objects over a wide range of systems (from cataclysmic variables to super massive BH in AGN). 
\begin{figure}
\begin{center}
\begin{tabular}{c}
\includegraphics[height=7cm]{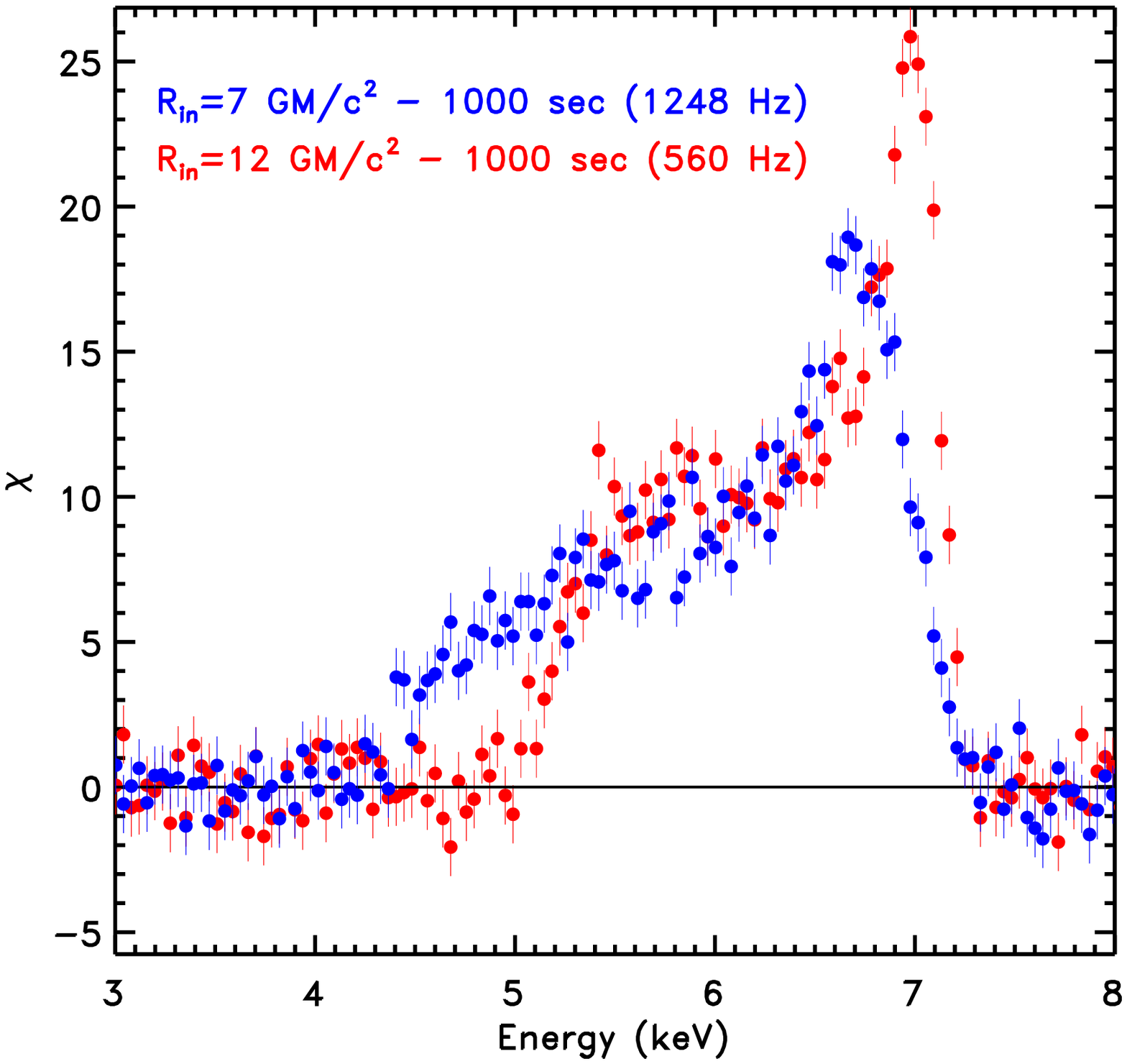}\includegraphics[height=7cm]{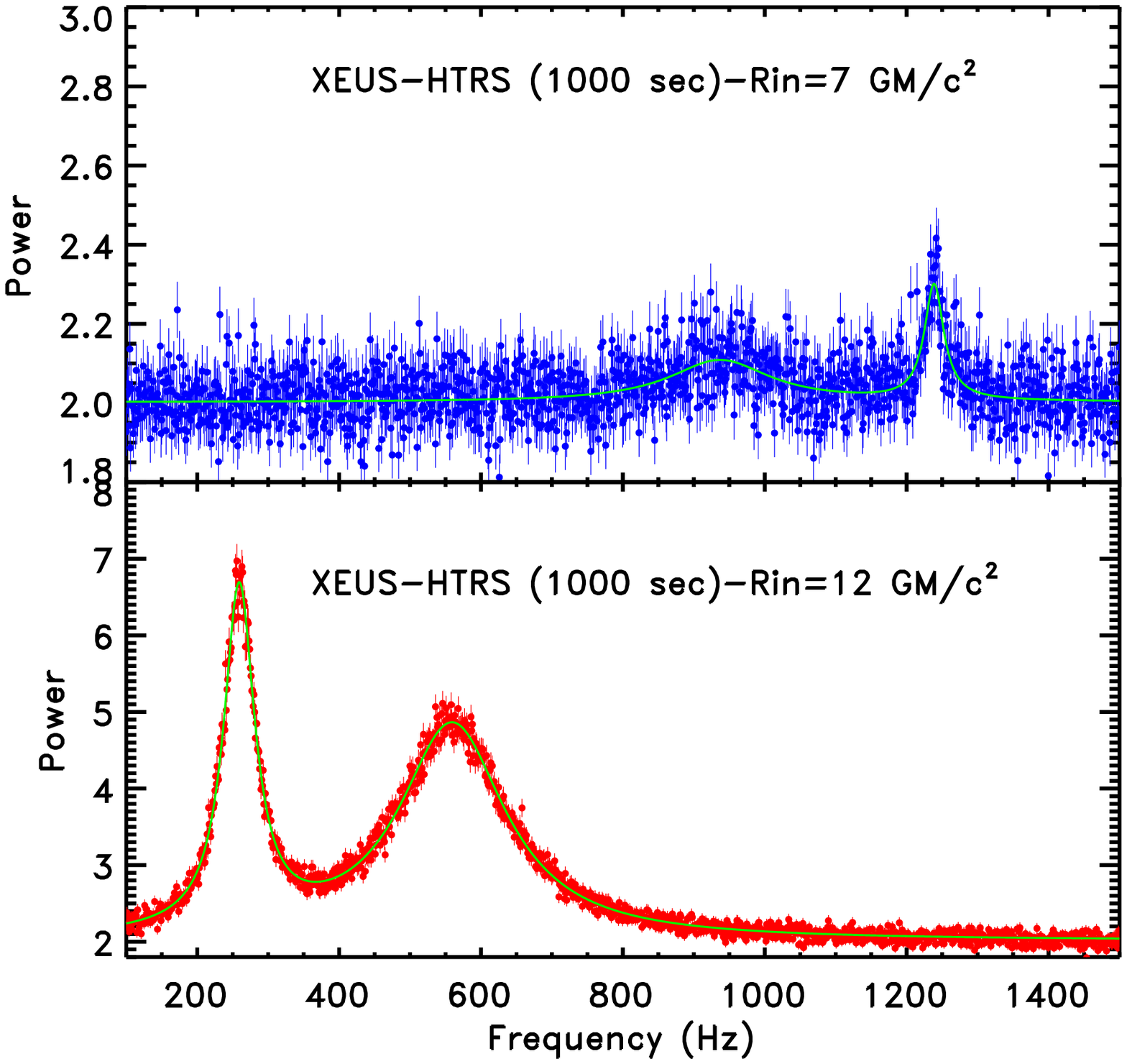}
\end{tabular}
\end{center}
\caption[example] 
{ \label{fig:linesandpds} 
Left) The residuals ($\chi^2$ deviations) for two simulated relativistically broadened Iron lines at two different inner disk radii (R$_{\rm in}$=$\rm 7GM/c^2$ and R$_{\rm in}$=$\rm 12 GM/c^2$). The continuum and line parameters are taken from Cackett et al.\cite{cackett}. Right) Simulated Power Density Spectra, assuming that the upper QPO is a Keplerian frequency at the same inner disk radii, corresponding to frequencies of 1248 Hz and 560 Hz respectively. On 1000 seconds, both diagnostics can be obtained from a joint spectral and timing analysis of the same data set.}
   \end{figure} 

To illustrate the complementarity between timing and spectral studies, in Fig. \ref{fig:linesandpds}, we show the energy spectra simulated for 1000 seconds from a $\sim$ 150 mCrab X-ray source displaying a relativistic Iron line, with parameters for the continuum and the line taken from\cite{cackett}. Two inner disk radii are assumed in the simulations: R$_{\rm in}$=7GM/c$^2$ and 12 GM/c$^2$. As can be seen, even on this timescale, the two profiles can be easily distinguished; the one produced at smaller radius (i.e. closer to the innermost stable circular orbit) being broader than the one produced at larger radius. Now assuming that the upper kHz QPO has a Keplerian orbital frequency at R$_{\rm in}$, one can simulate a Fourier Power Density Spectrum (PDS) with the corresponding QPOs over the same timescale. For the sake of simplicity, we have deduced the frequency of the lower kHz QPOs from the upper one by subtracting 300 Hz. The parameters of both QPOs (i.e. width and RMS amplitudes) have been taken as representative values of NS QPOs at those frequencies\cite{barret}. As can be seen, the two QPOs are clearly detected, including the lower QPO whose coherence and amplitude drops at high frequency (with the possibility that this effect might be related to the ISCO\cite{barret07}). Simultaneous measurements of the inner disk radius from spectral fitting and from kHz QPOs will therefore enable us to test the hypothesis that the upper QPOs is orbital. If so, constraints on the mass and radius of the NS will be derived (and so for a large sample of systems). Furthermore, if the ISCO signature is confirmed, a direct measurement of the NS mass will be obtained, together with a constraint on its radius, thus providing a direct constraint on the equation of state of dense matter.

The ability to detect both NS and BH QPOs on timescales comparable to their coherence timescales, defined as $1/\pi\Delta\nu$, ($\Delta\nu$ is the FWHM of the QPO recovered from a Lorentzian fit) is demonstrated in Fig. \ref{fig:cohedet}, where we show for various $\Delta\nu$, the required RMS amplitude for a 5$\sigma$ detection. The regions, in which NS and BH QPOs fall are indicated with red hashed rectangles\cite{rm,vdk}. As can be seen, the 67 Hz QPO detected from GRS1915+105, whose width reaches 3 Hz and its amplitude is of the order of 1\% in the 2--30 keV band, when its intensity ranges from 1 to 2 Crab\cite{morgan} will be detected on its coherence timescale. For NS, the same conclusions is reached for most sources, for both the narrower and weaker amplitude lower kHz QPOs, and the broader, higher amplitude upper kHz QPOs\cite{vdk}.

\subsection{Matter at supra-nuclear densities}
In NS the density in the core can be several times nuclear. Similarly, the magnetic fields of NS can exceed by ten orders of magnitude the strongest fields generated in terrestrial laboratories. Hyperon-dominated matter, deconfined quark matter, superfluidity, even superconductivity are predicted in NS. Similarly, quantum electrodynamics predicts that in strong magnetic fields the vacuum becomes birefringent. This makes NS ideal laboratories, not only for astrophysics, but also for nuclear and particle physics\cite{lattimer,hardinglai}. 

Different Equations of State (EoS) of dense matter predict different maximum masses and mass-radius relations. Determining the EoS requires measuring the mass (M) and radius (R) of the NS simultaneously. So far the most accurate mass measurements have been obtained for binary radio pulsars, all measures clustering around the canonical 1.4 M$\odot$ a value which can be accommodated by virtually all EoS. The HTRS, by its ability to cope with 2 millions events per second, will probe NS radii and masses, and determine the physical state of matter in its densest form found in the observable Universe. 

A statistically significant sample of NS will be observable over a wide range of luminosities, from the dimmest states up to the brightest states during Eddington limited X-ray bursts, over a wide range of ages (from birth to $10^{10}$ years), in a variety of conditions, with the NS being powered either by accretion, nuclear energy, internal heat release, or magnetic energy. X-rays alone provide several complementary diagnostics for the same object. For the EoS, these diagnostics include i) X-ray burst spectroscopy, enabling us to detect gravitationally redshifted emission lines and absorption edges, ii) waveform fitting of X-ray pulsations produced by rotating hot spots, either during X-ray bursts or in the persistent emission of pulsars, iii) X-ray spectroscopy of cooling NS and measurements of the associated cooling curves whose shape depends on the NS structure and internal composition, iv) the study of the sub-ms variability from the innermost regions of accretion disks, and v) the detection of QPOs due to seismic 
vibrations in magnetars after giant flares. The count rate capability of the WFI and NFI (which have both improved energy resolution compared to the HTRS) will not be sufficient to observe the brightest phases of NS. On the other hand, both instruments will be very useful to observe them in their dimmest states (e.g., cooling NS) and to observe similar accreting BH and NS in other galaxies. Note however that accretion supplies metals to the atmosphere, hence makes more promising the search for redshifted spectral lines in bright accreting systems, e.g. during type I X-ray bursts.

While obtaining such diagnostics, nuclear burning will also be probed and the properties of NS atmospheres, which depend on the magnetic field will be measured. Similarly XEUS will estimate the spin frequency distribution of NS spun up by accretion and test the exciting hypothesis that the distribution is bound below 750 Hz, due to angular momentum losses via magnetic torques or gravitational radiation\cite{chak}. For highly magnetized systems (e.g. magnetars), their extreme magnetic fields will be measured directly with X-ray spectroscopy through the unambiguous detection and identification of cyclotron resonance scattering features\cite{hardinglai}. 

\begin{figure}
\begin{center}
\begin{tabular}{c}
\includegraphics[height=7cm]{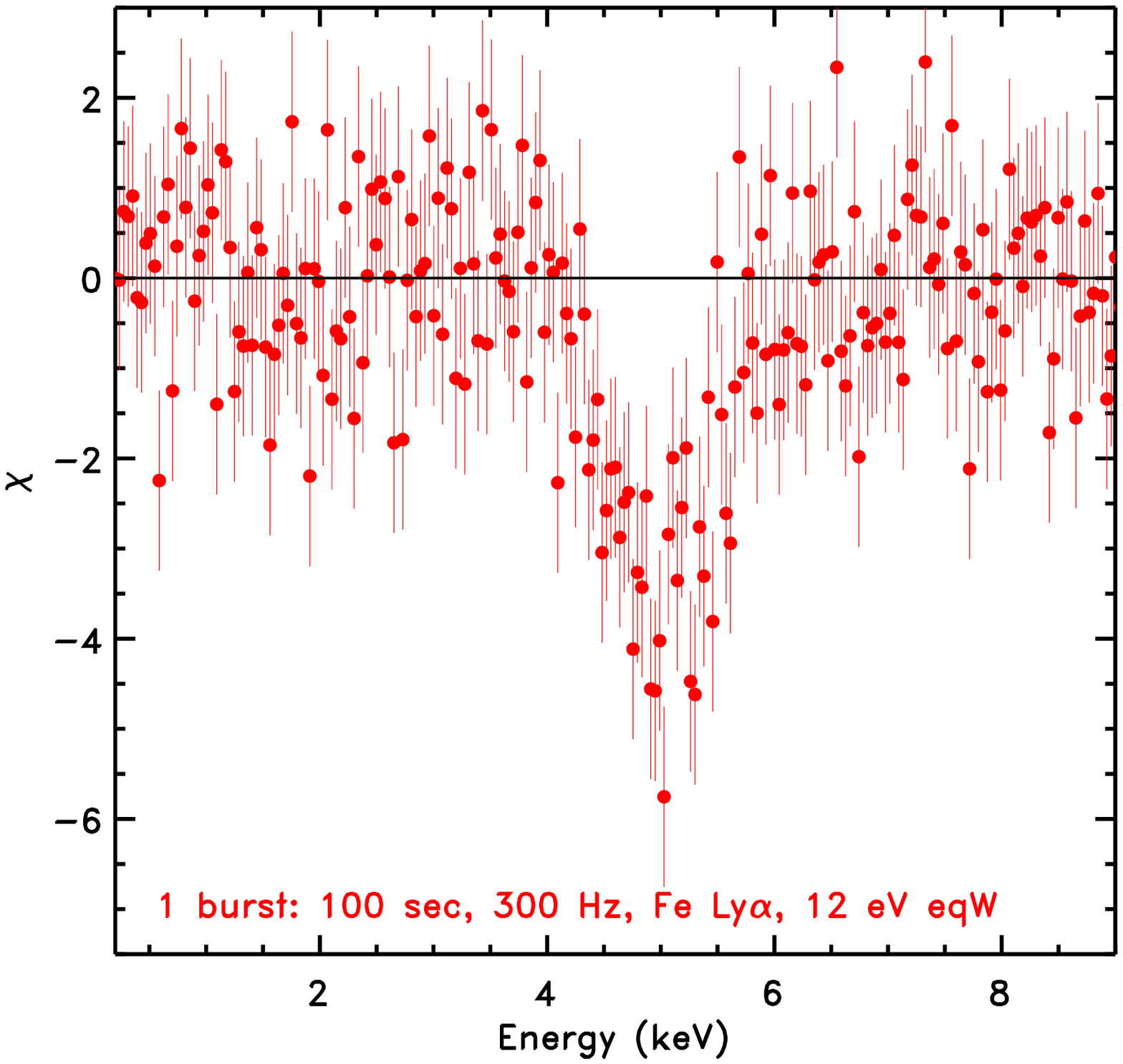}\includegraphics[height=7cm]{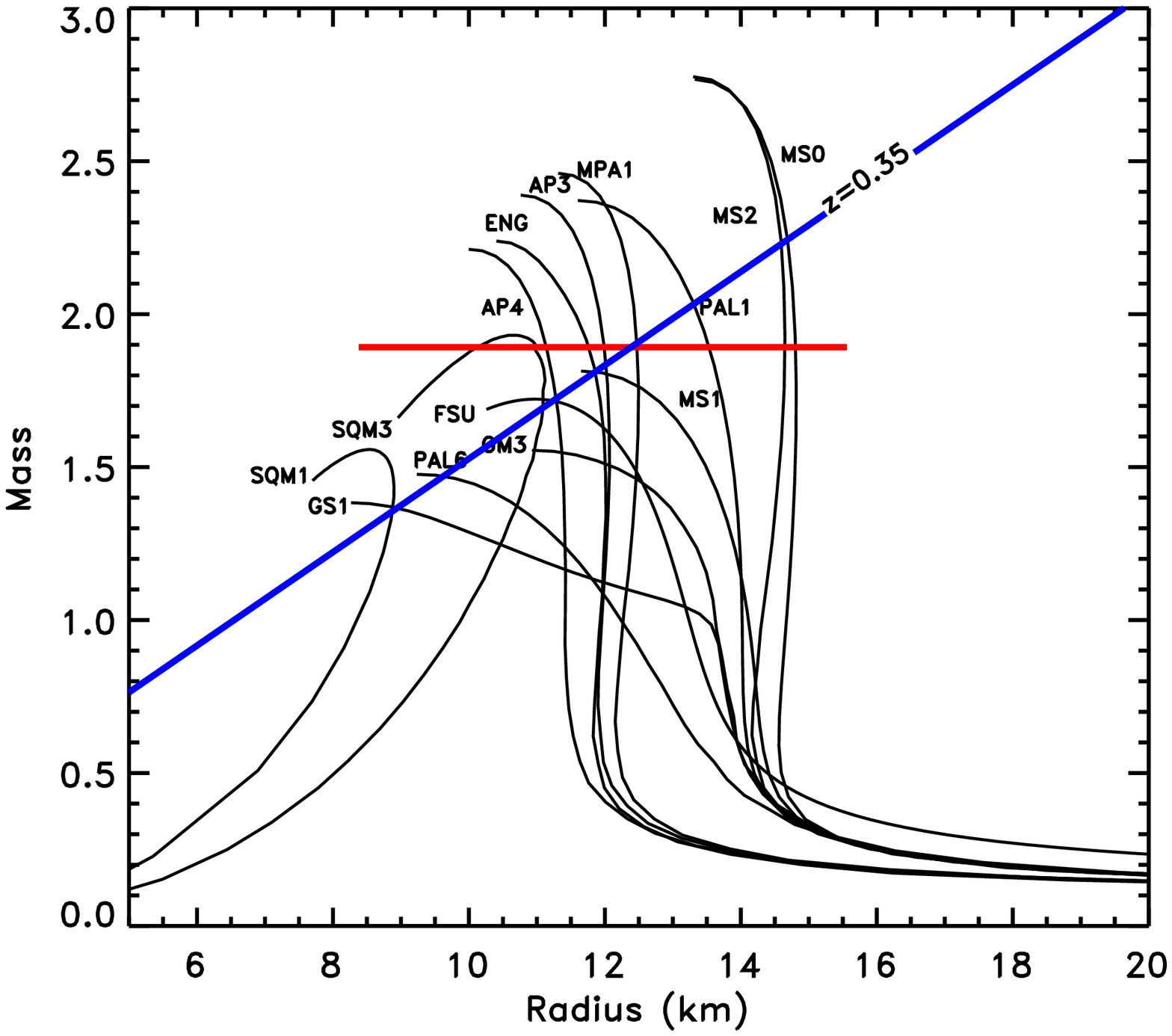}
\end{tabular}
\end{center}
\caption[example] 
{ \label{fig:lines} 
Left) The residuals ($\chi^2$ deviations) for a simulated redshifted (z=0.35) Fe Ly$\alpha$ absorption lines, predicted in type I X-ray bursts. The equivalent width of the line is set at 12 eV\cite{chang2005}, and the duration of the observation is 100 seconds.  The burst continuum spectrum is assumed to be a blackbody of 1.7 keV and an X-ray flux of  $4\times 10^{-8}$ ergs/s/cm$^2$. The width of the line includes contributions from magnetic (Zeeman or Paschen-Back) splitting by the starÕs magnetic field, longitudinal and transverse Doppler shifts, special relativistic beaming, gravitational redshifts, light bending, and frame dragging\cite{chang2005, bhat2006}. It is approximated here by a gaussian absorption line of $\sigma\sim 450$ eV, as expected for a NS of 300 Hz spin frequency (values courtesy of S. Bhattacharyya). Right) Various mass-radius relations for different equations of state of dense matter (data are courtesy of J. Lattimer). The constraints inferred from a redshift and the identification of an orbital frequency at the ISCO (through $\rm M_*\approx 2.2\,M_\odot\times(1000~{\rm Hz}/\nu_{\rm ISCO})$) are shown as continuous lines in blue and red (the lower limit on the radius is set by the causality constraint). This illustrates that just by two such measurements, the correct equation of state could be identified. }
\end{figure} 

As illustrative examples of the breakthroughs expected with XEUS, we show the constraints on M and R set by the detection of weak redshifted Iron line expected during type I X-ray bursts\cite{chang2005, bhat2006} in Fig. \ref{fig:lines}. A correct identification of the line would enable the redshift (z) to be measured, hence a M/R ratio, through $\rm M/R=\frac{c^2}{2G}(1-(1+z)^{-2})$. Combining this diagnostic, with for instance, the identification of an orbital frequency at the innermost stable circular orbit would pinpoint directly the correct equation of state of dense matter. With current instrumentation, such measurements are impossible, and not only because of reduced sensitivity compared to the HTRS: XMM-Newton and Chandra are unable to cope with count rates typical of X-ray bursts, whereas the RXTE/PCA has inadequate spectral resolution. 
\section{THE HIGH TIME RESOLUTION SPECTROMETER}
\begin{figure}
\begin{center}
\begin{tabular}{c}
\includegraphics[height=4cm]{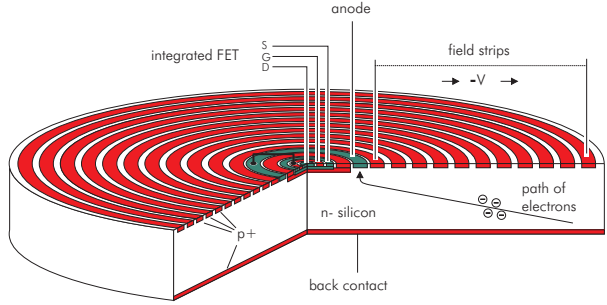}
\end{tabular}
\end{center}
\caption[example] 
{ \label{fig:sdd} 
Schematic cross section of a cylindrical Silicon Drift
Detector (SDD). Electrons are guided by an electric field towards
the small collecting anode located at the center of the device.
The first transistor of the amplifying electronics is integrated
on the detector chip.
}
   \end{figure}

In its current design, the first stage of the HTRS is a compact array of 37 Silicon Drift Detectors (SDDs) operated out of focus of the XEUS optics. Among the fast X-ray detectors currently available, SDDs are the most promising. The SDD is a completely depleted volume of silicon in which an arrangement of increasingly negatively biased rings drive the electrons generated by the impact of ionising radiation towards a small readout node in the centre of the device (see Fig. \ref{fig:sdd}). The time needed for the electrons to drift is much less than 1$\mu$s. The main advantage of SDDs over conventional PIN diodes is their small physical size and consequently the small capacitance of the anode, which translates to a capability to handle very high count rates simultaneously with good energy resolution\cite{lechner1996,lechner2000,strueder2000}. To take full advantage of the small capacitance, the first transistor of the amplifying electronics is integrated onto the detector chip. The stray capacitance of the interconnection between the detector and amplifier is thus minimized and, furthermore, the system becomes practically insensitive to mechanical vibrations and electronic pickup. 

\begin{figure}
\begin{center}
\begin{tabular}{c}
\includegraphics[height=7cm]{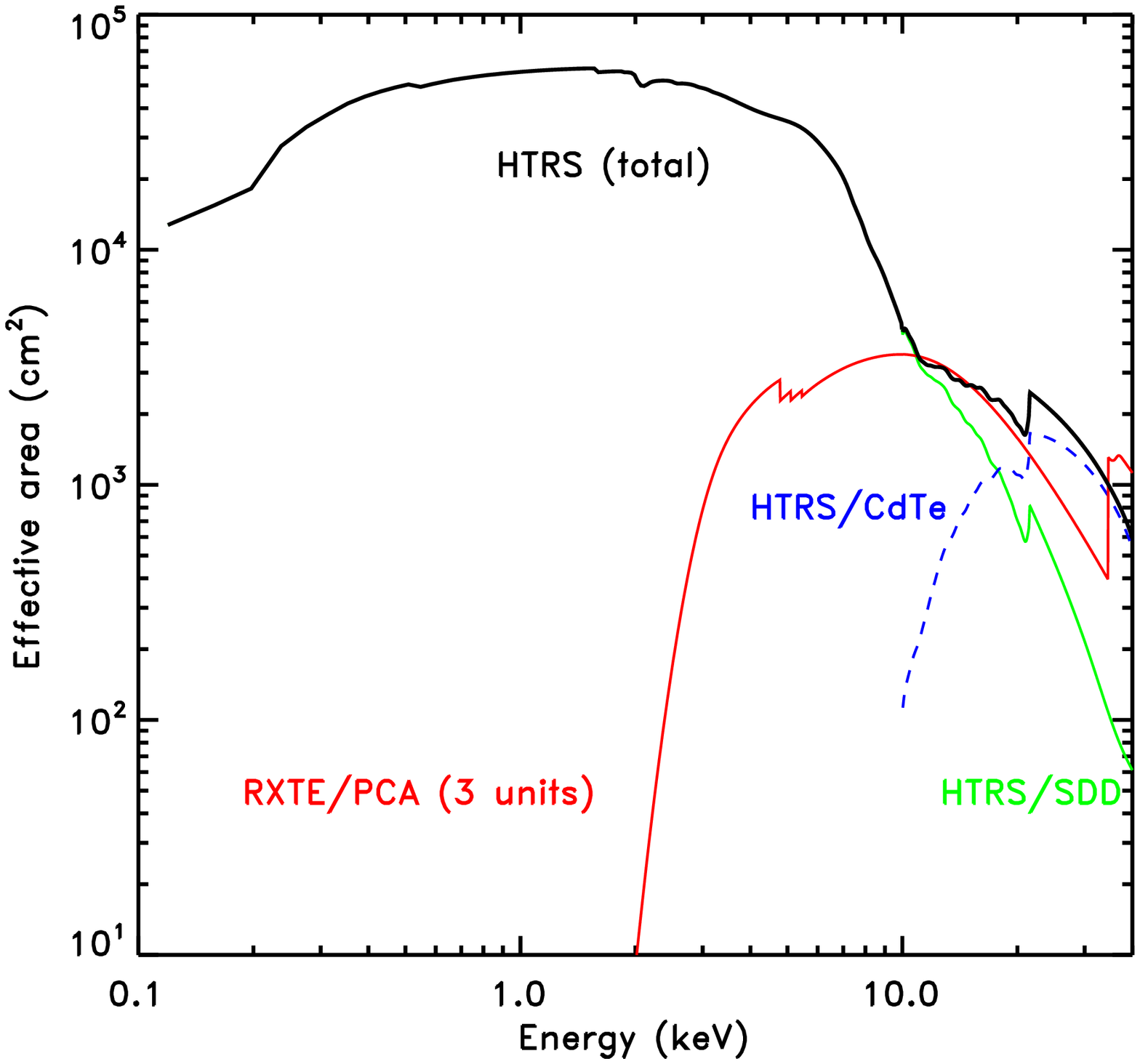}\includegraphics[height=7cm]{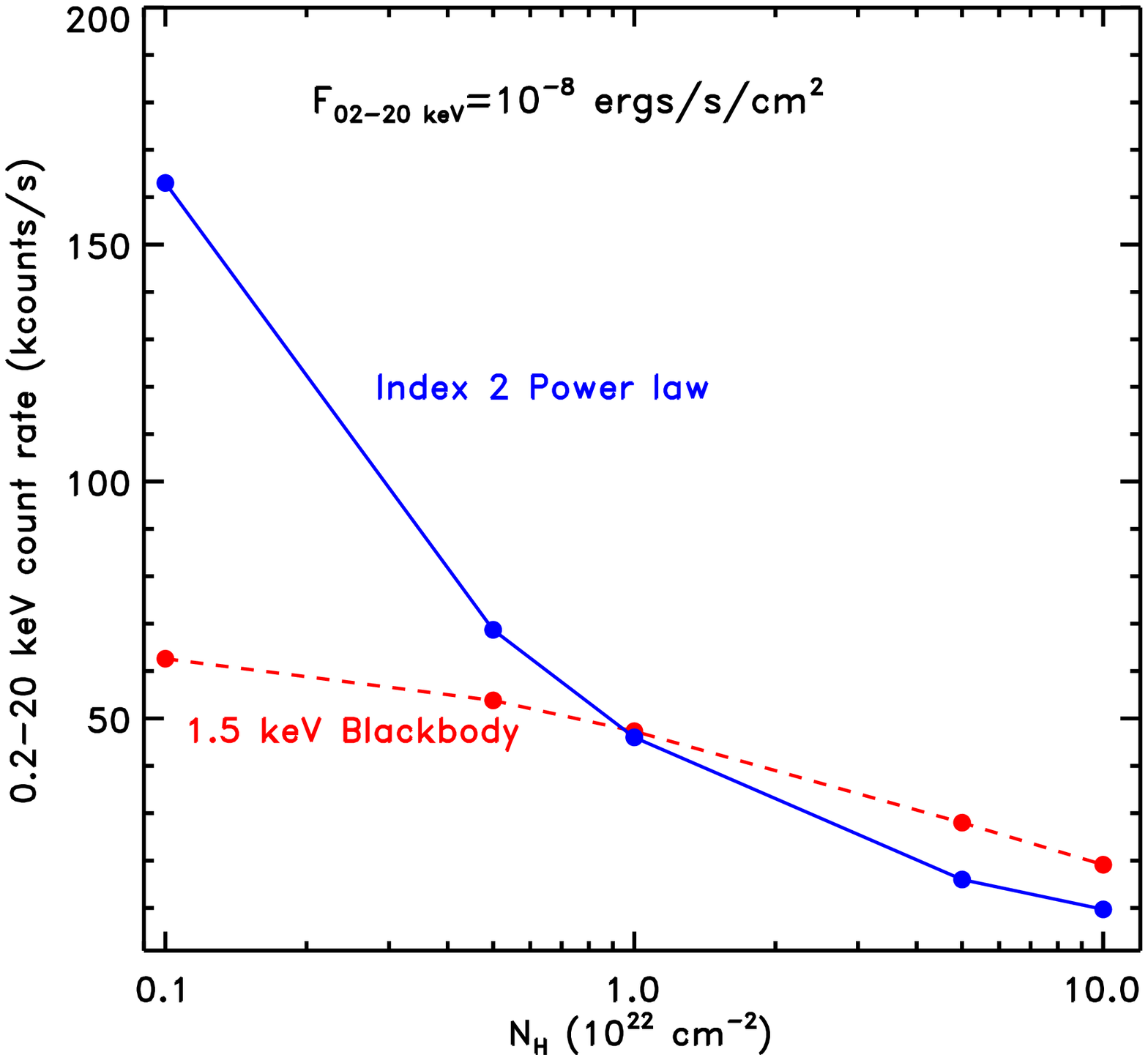}
\end{tabular}
\end{center}
\caption[example] 
{ \label{fig:effcurve} 
Left) The effective area of the HTRS plotted as the sum of the effective area in the SDD stage and the effective area in the CdTe detector located underneath the SDD array. For comparison the effective area of the RXTE/PCA is shown for three out of the five units of the PCA (with which a majority of observations was performed). This shows that the HTRS has an effective area about twice larger than the PCA at 20 keV.  Right) Expected count rates in kcounts/s for various column density (N$\rm_H$) and for two different spectral shapes, normalized to the same unabsorbed flux of $10^{-8}$ ergs/s/cm$^2$ in the 0.2-20 keV band (the Crab is $6.8 \times 10^{-8}$ ergs/s/cm$^2$ in the same energy band, assuming a power law spectrum as N(E)=9.7 E$^{-2.1}$ photons/cm$^2$/s/keV).}
   \end{figure}

The detection efficiency of the 450 $\mu$m thickness of the SDD decreases rapidly above 10 keV. In order to cope with the high-energy response of the mirrors, as foreseen for XEUS, a higher Z semiconductor detector must be placed under the Silicon. This could be achieved using a pixilated array of Cadmium Telluride of 1 mm thickness, of the same size of the SDD array (in a similar way to the Hard X-ray Camera below the Wide Field Imager). Such detectors are currently being developed for the ECLAIRs instrument on-board the SVOM mission\cite{ehanno}. The count rate expected in the CdTe detector is typically 600 counts/s above 10 keV for a Crab like spectrum. The effective area of the combined stages of the HTRS detector is shown in Fig. \ref{fig:effcurve}, together with the expected count rates for two different spectral shapes.

The instrument specifications derive from the requirement for the HTRS to observe the brightest X-ray sources in the sky, mostly X-ray binaries, either transient or persistent, in their quiet or burst phases. Simulations using the current effective area of the mirrors show that the Crab would produce about 457 kcounts/s (above 0.2 keV). The requirement is therefore to be capable of observing, a source of $\sim 5$ times the intensity of the crab (this encompasses the intensity of a wide majority of currently known potential targets for the HTRS). In addition, the arrival time of each photon should be recorded with a timing resolution of about 10$\mu$s, the energy resolution of the detector should be better than 200 eV.

\begin{figure}
\begin{center}
\begin{tabular}{c}
\includegraphics[height=7cm]{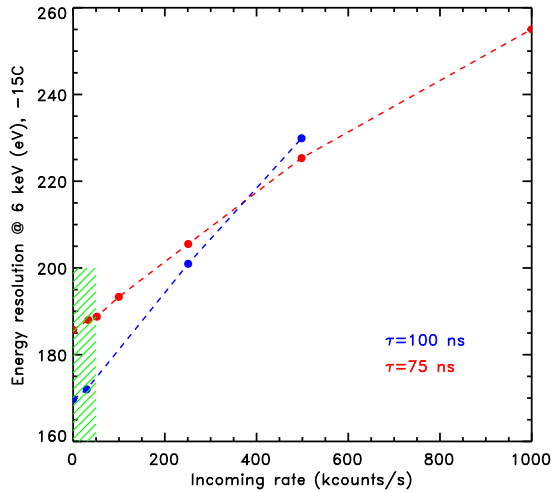}
\end{tabular}
\end{center}
\caption[example] 
{ \label{fig:reseneversusrate} 
Energy resolution of a single SDD as a function of the incoming rate of photons for two shaping times (75 ns in red  and 100 ns in blue). Thanks to the defocusing of the HTRS, the region of operation of the SDD will be limited to the green hashed region. This figure shows that an energy resolution as good as 190 eV is possible (the data are courtesy of P. Lechner). }
   \end{figure}

For timing studies, dead time is always a critical issue. Dead time will include contributions from the signal rise time, the charge-sensitive pre-amplifier and the shaping amplifier. The first two of these can be very short, and the limiting contribution is that of the shaping amplifier, where a trade-off between speed and energy resolution is necessary. Shaping time constants as short as 75 ns have been found to be usable (see Fig.\ref{fig:reseneversusrate}). This translates to a minimum feasible dead time of about 300 ns. Using currently available devices and pipelining techniques, the analogue-digital conversion stage is not a limiting factor at these speeds. For the analog electronic chain, a 300 ns dead time per event (which is the time to process an event) corresponds to a 3\% dead time for a source producing 100 kcounts/s per individual SDD. The analog electronics, as well as the overall design of the HTRS is currently being studied at CESR. The option that the analog electronics is replaced by a digital electronics (after the pre-amplifier) is also being considered at CESR, as part of an R\&D program currently funded by the French Space Agency (CNES).  Fig. \ref{fig:reseneversusrate} shows that the energy resolution of the SDD for a shaping time constant of 75 ns, at count rates below 50 kcounts/s is much less than 200 eV, hence demonstrating a high technological readiness level for the HTRS.
\section{CONCLUSIONS}
XEUS is currently in its assessment study phase at ESA, along which all aspects related to the mission will be carefully evaluated. The electrical, thermal, mechanical design (including filter wheel, baffle) of the HTRS is being optimized as part of this study. No major issues have been identified so far, indicating that the HTRS will have capabilities matching the science requirement of the mission to observe bright X-ray sources to probe strong gravity and dense matter using X-rays generated in the vicinity of black holes and neutron stars.

\acknowledgments
DB is pleased to thank the HTRS project team at CESR: C. Amoros, K. Lacombe, J. Land\'e, P. Mandrou, D. Rambaud, R. Pons. DB is grateful to L. Str{\"u}der, P. Lechner, A. Niculae, H. Soltau (MPE/MPI and PN-Sensors) for sharing their expertise in Silicon Drift Detectors, and for providing us with inputs for the HTRS definition. Special thanks to Andy Fabian for stimulating discussions. Finally DB wishes to thank all the members of the XEUS Study Science Team for their efforts in getting XEUS as the best possible X-ray astrophysics mission for the future.

\bibliography{barret-vf}   
\bibliographystyle{spiebib}   

\end{document}